\documentclass[pra,twocolumn,showpacs,preprintnumbers,superscriptaddress]{revtex4}

\usepackage{color}
\usepackage{times}
\usepackage{subfig}
\usepackage{bm}
\usepackage{graphicx}
\usepackage{amsbsy}
\usepackage{amsmath}
\usepackage{amsfonts}
\usepackage{amsthm}
\usepackage{float}
\usepackage{amssymb}
\DeclareMathOperator{\Tr}{Tr}
\definecolor{purple(html/css)}{rgb}{0.5, 0.0, 0.5}
\newcommand{\ket}[1]{| #1 \rangle}
\newcommand{\bra}[1]{\langle #1 |}
\begin{document}

\theoremstyle{plain}
\newtheorem{theorem}{Theorem}
\newtheorem{lemma}[theorem]{Lemma}
\newtheorem{corollary}[theorem]{Corollary}
\newtheorem{proposition}[theorem]{Proposition}
\newtheorem{conjecture}[theorem]{Conjecture}

\theoremstyle{definition}
\newtheorem{definition}[theorem]{Definition}

\title{\bf Complementarity of Quantum Correlations in Cloning and Deleting of Quantum State }
\author { Sk Sazim}
\affiliation{Institute of Physics, Sainik School Post, 
Bhubaneswar-751005, Odisha, India}
\author {Indranil Chakrabarty  }
\affiliation{International Institute of Information Technology, Gachibowli, Hyderabad 500 032, Andhra Pradesh, India.
}
\author {Annwesha Datta }
\affiliation{ Department of Physics, Indian Institute of Technology, Kanpur 208016, India.}
\author {Arun K. Pati}
\affiliation{ Harish-Chandra Research Institute, Chhatnag Road, Jhunsi, Allahabad 211 019, India}
\date{\today}

\begin{abstract}
We quantify the amount of correlations generated between two different 
output modes in the process of imperfect cloning and deletion processes. We use three different measures of quantum correlations and investigate their role in determining the fidelity of the cloning and the deletion. We obtain a 
bound  on the total correlation generated in the successive processes of cloning and deleting operations. 
This bound displays a new kind of complementary relationship between the quantum correlations 
required in generating a copy of a quantum state and the amount of correlations required in bringing it 
back to the original state by deleting and vice versa. 
Our result shows that the better we clone (delete) a state, the more difficult it will be 
to bring the state back to its original form by the process of deleting (cloning). 
\end{abstract}

\maketitle

\section{Introduction }
In the last two decades quantum information processing has emerged as a powerful tool
for implementing several tasks that cannot be done using classical means. Information
processing tasks such as super dense coding \cite{dense}, 
teleportation \cite{teleport}, remote state preparation \cite{remote} and key generation 
\cite{crypto}  are no longer only theoretical possibilities but also have been experimentally demonstrated.  
In all these protocols, quantum entanglement or quantum correlations in a broader sense, plays a pivotal role in 
making these information processing tasks successful.

Quantifying the amount of quantum correlations present in a pure bipartite system  is straightforward since this is 
given by the amount of entanglement present in the system. There are certain standard measures, 
more specifically entanglement monotones, quantifying the amount of entanglement for both pure as well as mixed 
states. The two computable measures are negativity \cite{vid} and concurrence \cite{SH,WW}. 
 However, there are certain open issues in understanding 
the nature of correlations present in a mixed state, or a multiqubit state.
It is not evident whether all the information-processing tasks that can be done more efficiently with 
quantum systems require entanglement as a resource. In the past, it has been
shown that even in the absence or near absence of entanglement one can carry out 
some information-processing tasks more efficiently in the quantum world \cite{lafff}.
Therefore, it is legitimate to ask, if not entanglement then what is responsible for such a behavior.
It has been suggested that the amount of correlations present in a composite system is not entirely captured by  entanglement.
The quantity which captures quantum correlations and gives a meaningful explanation of such behavior 
is quantum discord \cite{oli,hen,luo,ur,ani,mw10,ind1,horo}. This  aims to capture the non-classical correlations present in a system, 
and those that cannot be witnessed with entanglement. In addition to these standard approaches to quantifying 
the correlations in quantum mechanical systems, there are other approaches of quantifying quantum correlations. 
The most important of these are geometric measures \cite{dak,ade}.

In  quantum information theory the no-cloning theorem plays a fundamental 
role \cite{woo,adri,yuen}. This theorem states how nature prevents us from amplifying an unknown quantum state. However, 
in principle it is always possible to construct a quantum cloning machine that replicates an unknown quantum 
state approximately \cite{buz,mas,mass,buz1,dua,ind2}. These approximate quantum cloning machines can be of two types.
One is a state-dependent quantum cloning machine, for example, the Wootters-Zurek (WZ) quantum cloning machine, whose
copying quality depends on the input state \cite{woo}. The other type is a universal
quantum copying machine, for example, the Buzek-Hillery (BH)
quantum cloning machine \cite{buz}, whose copying quality remains the 
same for all  input states.
In addition, the performance of the universal BH quantum cloning machine is, on the average, 
better than that of the state-dependent WZ cloning machine. The fidelity of cloning of the BH universal quantum 
copying machine is $\frac{5}{6}$ - the optimal fidelity for the universal quantum cloning 
machines. Although it is impossible to copy a state perfectly, one can probabilistically  clone a quantum state, secretly chosen from a certain set 
of linearly independent states \cite{dua}. Also, it is possible to have linear superposition of multiple 
clones and obtain a probabilistic cloning machine as a special case of the former \cite{akp}.
Quantum deletion \cite{pati} on the other hand, is about the impossibility of deleting an 
arbitrary quantum state. More specifically, it states that the linearity of quantum theory precludes deleting an
unknown quantum state from two identical copies in either a 
reversible or an irreversible manner. The principle behind  quantum deletion will be clearer, if we 
compare the deletion operation with the Landauer erasure operation \cite{land}. Erasure of classical or quantum information 
cannot be performed reversibly. The erasure principle says that a single copy of 
some classical information can be erased at the cost of some energy. Thermodynamically, it is an irreversible operation. 
In quantum theory the erasure of a single unknown state is considered as swapping it with some standard state 
and then trashing it into the environment. In contrast, quantum deletion \cite{pati} is more of reversible ’uncopying’ of an unknown quantum state. 
It has been shown that in addition to the linear structure of quantum mechanics, other principles like unitarity, 
nosignalling, incomparability and conservation of entanglement are not congruous  to the concept of 
perfect deletion \cite{ Horodecki03, Horodecki05,
Nielsen99, pati03, Bhar07}. However, if one tries to delete an unknown 
quantum state probabilistically, then it is possible with a success probability of less than unity \cite{feng02}. It has also been 
shown that using these probabilistic deletion machines one cannot send superluminal signals probabilistically \cite{chakrabarty06}. 
Since perfect deletion is not possible, it is interesting to see whether 
one can delete an unknown state imperfectly. Researchers have devised various approximate deletion machines. 
These deletion machines are either state dependent or state independent \cite{pati100, adhikari04, adhikari05, adhikari06, qiu02, qiu102}. 
Recent explorations have revealed that one can construct a universal quantum deletion machine \cite{adhikari05}, and its fidelity can 
be further enhanced by the application of suitable unitary transformation \cite{adhikari06}. These deletion machines can 
have various applications in quantum information theory \cite{Horodecki04,Chakrabarty10,Srikanth07}. However, the optimal quantum deletion machine has not been found yet.

At this point one might ask an important question whether quantum correlations are responsible for our inability to 
produce high fidelity states in the approximate cloning 
or deleting a quantum state? Note, that initially there are no correlations between the input states. This is because 
they are the individual systems which are in a product state. However, at the output port we always obtain a combined state, which is  
usually correlated. A priori, it is not clear whether this correlations play an important role in deciding the fidelity of 
cloning and in deleting an unknown quantum state. 
In order to find an answer to this question, we consider a particular type of cloning machine, 
the BH cloning machine, and try to quantify the amount of correlations present in the mixed two qubit 
output state.  Similarly, for the deletion operation we consider a state-dependent quantum deleting machine to find out the 
correlations in the output modes. The basic motivation is to see how the correlations regulates the fidelity of 
the cloning and deletion processes. We find that the more the output modes are 
correlated the less is the fidelity in either cases. In other words, the process of cloning and deletion will be  more 
perfect if the output modes are poorly correlated. We quantify these correlations 
with three different kinds of measures and make this observation more precise.
The problem of complementarity or mutually exclusive aspects of quantum phenomena arose with the birth of 
quantum mechanics, soon after, Heisenberg discovered the uncertainty principle for the momentum and the position \cite{dua1}. 
A year later, Bohr proposed the concept of complementarity \cite{nbr}. Even in the domain of quantum information theory, 
the idea of complementarity is not new, as some authors have shown that there does exist the complementarity between 
the local and non-local information of  quantum systems \cite{oppen}.     
In this work we  observe a new kind of complementarity  in terms of successive correlations generated in 
the system when a state undergoes deletion after the cloning or the cloning after the deletion.
 
The organization of our paper is as follows. In section II, we provide a short introduction to the relevant measures 
of quantum correlations. In section III, we analyze the correlations content of the output of the Buzek-Hillery 
quantum cloning machine. We also analyze how the correlations content of the output modes plays a pivotal role in 
determining the fidelity of cloning. In section IV, we study the standard approximate deleting machine to obtain 
a correspondence between the fidelity of deletion and the amount of correlations generated in the process. 
In section V, we obtain a new kind of complementarity relation between the correlations generated in the system for  
the process of successive cloning and deletion, and also for the case when we clone the state after deletion. This complementarity 
gives a new bounds to the total correlations generated in the context of quantum correlation measures.  
\section{Various measures of Quantum Correlations}
For the sake of completeness, in this section, we give a brief description of three different measures 
which we will use to quantify the quantum correlations generated in  
cloning and deletion operations. These measures are (i)  negativity, (ii)  quantum discord and (iii) geometric Discord. 
Each of them 
represents three different classes of measures. We would like to see how generic  the 
complementarity is for cloning and deleting if we use different measures of quantum correlations.
\subsection{Measures in entanglement-separability paradigm: Negativity}
Negativity is an entanglement monotone that quantifies how strongly the partial transpose of a density operator 
fails to become positive \cite{vid}. 
The negativity, \(N(\rho_{AB})\), of a bipartite state \(\rho_{AB}\) is defined as the absolute value of the sum of 
the negative eigenvalues of \(\rho_{AB}^{T_{A}}\). Alternatively, we can find negativity by the following relation
\begin{equation}
 N(\rho_{AB})=\frac{||\rho_{AB}^{T_{A}}||_1-1}{2},
\end{equation}
where $||A||_1$ is called trace-norm of $A$ and it is defined as $||A||_1= {\rm Tr}[\sqrt{A^{\dagger}A}]$. 
Here, \(\rho_{AB}^{T_{A}}\) denotes the partial transpose of \(\rho_{AB}\) with respect to the subsystem \(A\). 
If we have a general state $\rho_{AB} = \sum_{ijkl} p^{ij}_{kl} |i\rangle \langle j | \otimes |k\rangle \langle l| $,
its partial transpose (with respect to the $A$ party) is defined as
$\rho_{AB}^{T_A} = \sum_{ijkl} p^{ij} _{kl} (|i\rangle \langle j | )^T\otimes |k\rangle \langle l| = \sum_{ijkl} p^{ij} _{kl} |j\rangle \langle i | \otimes |k\rangle \langle l|$ \cite{peres,Peres_Horodecki}. 
The logarithmic negativity is then defined in terms of $N(\rho_{AB})$ as
\begin{equation}
  E_{N}(\rho_{AB}) = \log_2 [2 N(\rho_{AB}) + 1].
\label{eq:LN}
\end{equation}
For two-qubit states, 
\(\rho_{AB}^{T_{A}}\) has at most one negative eigenvalue \cite{rana}. It has been seen that for two-qubit states, 
a positive logarithm negativity implies that the state is entangled and distillable, whereas 
a vanishing logarithm negativity implies that the state is separable \cite{peres, Peres_Horodecki}.
\subsection{Information theoretic measure: Quantum Discord}
Information theoretic measures are  constructed                                                                                                                             from the perspective of 
defining the notion of quantum correlations from the information theoretic viewpoint (entropic quantities). These measures are not 
computable in a closed form like the entanglement monotones. In spite of not being computable in a closed form, they 
can be efficiently computed numerically for two-qubit systems. 
The most important of them is quantum discord which shows that  quantum correlations in mixed states goes beyond the notion of entanglement.
Quantum discord is defined as the difference between two quantum information-theoretic quantities, whose classical 
counterparts give equivalent expressions for the classical mutual information. Quantum discord is nothing but 
the difference between the total correlations and the classical correlations present 
in the bipartite quantum systems, thus quantifying the amount of quantum correlations present in it. It is defined as 
\cite{oli,hen,luo} 
\begin{equation}
\label{eq:discord}
D(\rho_{AB})= {\cal I}(\rho_{AB}) - {\cal J}(\rho_{AB}).
\end{equation}
The total correlations, \({\cal I}(\rho_{AB})\), for a bipartite state \(\rho_{AB}\) is given by the 
mutual information
\begin{equation}
\label{qmi}
\mathcal{I}(\rho_{AB})= S(\rho_A)+ S(\rho_B)- S(\rho_{AB}),
\end{equation}
where $S(\varrho)= - {\rm Tr} (\varrho \log_2 \varrho)$ is the von Neumann entropy of the quantum state \(\varrho\), and 
 \(\rho_A\) and \(\rho_B\) are the reduced subsystems of the bipartite state \(\rho_{AB}\).
On the other hand, \({\cal J}(\rho_{AB})\) can be thought of as the amount of classical 
correlations in \(\rho_{AB}\), 
and is defined as \cite{hen} 
\begin{equation}
\label{eq:classical}
 {\cal J}(\rho_{AB}) = S(\rho_A) - S(\rho_{A|B}). 
\end{equation}
Here, $S(\rho_{A|B}) = \min_{\{\Pi_i\}} \sum_i p_i S(\rho_{A|i})$,
is the average entropy of the entropies of states $\rho_{A|i} $. The conditional entropy $S(\rho_{A|i})$  
is the entropy of the subsystem $A$ conditioned on a  measurement performed by \(B\) with a rank-one 
projection-valued measurement \(\{\Pi_i\}\).
These states are given by 
\(\rho_{A|i} = \frac{1}{p_i} {\rm Tr}_B[(\mathbb{I}_A \otimes \Pi_i) \rho (\mathbb{I}_A \otimes \Pi_i)]\), 
with probability \(p_i = {\rm Tr}_{AB}[(\mathbb{I}_A \otimes \Pi_i) \rho (\mathbb{I}_A \otimes \Pi_i)]\).
Here \(\mathbb{I}\) is the identity operator on the Hilbert space of \(A\). Note, that 
the discord is a positive quantity and vanishes for classical-classical and quantum-classical states.
\subsection{Geometric Measure: Geometric Discord}
Apart from these two classes of measures there is another way by which one can quantify the amount of 
quantum correlations present in a two qubit bipartite state. This is captured by the geometric measures of 
quantum correlations. It had been argued that the difficulty experienced in calculating quantum discord can be minimized, for a general 
two-qubit state, by defining its geometrical version \cite{dak}. It is well known that almost all 
(entangled or separable) states are disturbed by the measurement. However, there are certain states 
which are invariant under the measurement performed on the sub-system $A$. These states are the so called classical-quantum (CQ) states. 
A CQ density matrix is of the form
\begin{eqnarray}
\label{cq}
\rho = \sum _i  p_i  |i\rangle  \langle i | \otimes\rho _{i} ,
\end{eqnarray}
where $p_i$ is a probability distribution, $\{|i\rangle\}$ is 
an orthonormal set of vectors for $A$ and $\rho_{i}$ are the elements of $B$. 
A classical-quantum state is not affected by a measurement on $A$ in any case.
One can show that the state $\rho$ is of
zero-discord if and only if there exists a von Neumann measurement
$\{\Pi_k=|\psi_k\rangle\langle \psi_k|\}$ 
such that~\cite{ani}
\begin{equation}
\label{0dis}
\sum_k(\Pi_k\otimes \mathbb{I}_B)\rho(\Pi_k\otimes \mathbb{I}_B)=\rho.
\end{equation}
It had been seen in Ref. \cite{dak}, that these two states in equation (\ref{cq}) and (\ref{0dis}) are identical. 
Let $S$ be the set consisting of all classical--quantum two qubit states,  and let us assume that 
$\chi$ is a generic element of this  set. Then the geometric discord $DG$ of an arbitrary two-qubit state $\rho_{AB}$   
is given by the distance between  the state $\rho_{AB}$ and the closest classical-quantum state. Geometric discord 
has been introduced as
 \begin{eqnarray}\label{dgdef}
 DG(\rho_{AB})=2\min_{\chi \in S}||\rho_{AB} -\chi ||_2^2,
\end{eqnarray}
where the coefficient $2$ on the right hand side is the normalization factor and $||X -Y ||_2=\Tr(X-Y)^2$ is the 
square norm in the Hilbert-Schmidt space.   
For the geometric discord of the state $\rho_{AB}$ to have a nice closed form, one needs to express the state 
in terms of the Pauli matrices ($\sigma _1,\sigma _2,\sigma _3$) as
$\rho_{AB} = \frac 14(\mathbb{I}_{4}+\sum_{i=1}^3 x_i\sigma_i 
\otimes \mathbb{I}_{2} +\sum_{j=1}^3 y_j \mathbb{I}_{2}\otimes \sigma_j+\sum_{i,j=1}^3 t_{ij} \sigma _i\otimes\sigma_j)$,
where $t_{ij}=\text{Tr}[\rho(\sigma_i\otimes \sigma_j)]$, $\mathbb{I}_n$ is the 
identity matrix of order $n$, $\vec{x}=\{x_i\},\vec{y}=\{y_i\}$ represent the three-dimensional Bloch column vectors and $t=[t_{ij}]$ is the correlation matrix.  Then, we can rewrite the geometric discord 
as \cite{ade} $DG(\rho_{AB})= \frac 12(\|\vec x\|_2^2 + \|t\|_2^2 -4 k_{\max})=2 \text{Tr}[S]-2k_{\max}$,
with $k_{\max}$ being the largest eigenvalue of the matrix $S = \frac 14(\vec x {\vec  x}^{\sf T}+  t t^{\sf T})$ where '${\sf T}$' denotes transposition.
There are other approaches to define the geometric discord, however we focus only on
the above presented one.
\section{Analysis of the correlations content of the output of Buzek-Hillery copying machine}
In this section we consider the universal Buzek-Hillery cloning machine and quantify  the correlations present in 
the  output copies of the Buzek-Hillery cloning machine  \cite{buz}. 
But before that we give a short description of the Buzek-Hillery 
cloning machine.  We recall that the action of the Buzek-Hillery quantum cloning machine \cite{buz} is
given by the transformations
\begin{eqnarray}
|0\rangle_{a}|0\rangle_{b}|Q\rangle_{x}\longrightarrow|00\rangle_{ab}|Q_0\rangle_{x}
+[|01\rangle_{ab}+|10\rangle_{ab}]|Y_0\rangle_x,\nonumber\\
|1\rangle_{a}|0\rangle_{b}|Q\rangle_{x}\longrightarrow|11\rangle_{ab}|Q_1\rangle_{x}
+[|01\rangle_{ab}+|10\rangle_{ab}]|Y_1\rangle_x,\label{copytr}
\end{eqnarray}
where $a,b$ and $x$ denote qubits corresponding to input state port, blank state port and the machine state port.
The unitarity and the orthogonality of the cloning transformation demand the following conditions to be satisfied:
\begin{eqnarray}
\langle Q_i|Q_i\rangle_{x}+2\langle Y_i|Y_i\rangle_{x}&=&1~~~~(i=0,1),\nonumber\\
 \langle Y_0|Y_1\rangle_{x}=\langle Y_1|Y_0\rangle_{x}&=&0.
\end{eqnarray}
Here, we assume the  machine state vectors $|Y_i\rangle_{x}$ and
$|Q_i\rangle_{x}$ to be mutually orthogonal. This is also true for 
the state vectors $\{|Q_0\rangle,|Q_1\rangle \}$.\\
The unknown quantum state which is to be cloned is given by
\begin{eqnarray}
|\psi\rangle= \alpha|0\rangle+\beta|1\rangle,
\label{state}
\end{eqnarray}
where $\alpha,\beta$ are complex numbers satisfying, $|\alpha|^{2}+|\beta|^{2}=1$.  
After using the cloning transformation (\ref{copytr}) on the quantum state (\ref{state})
and then tracing out the machine state, the reduced
density operator describing the two qubit output modes of the original and the cloned state is given by
\begin{eqnarray}
\rho_{ab}^{clone}&=&(1-2\xi)(\alpha^2|00\rangle_{ab}\langle00|+\beta^2|11\rangle_{ab}\langle 11|){}\nonumber\\
&+&\frac{\alpha\beta}{\sqrt{2}}(1-2\xi)(|00\rangle_{ab}\langle \psi^+ | + 
|\psi^+\rangle_{ab}\langle 00|\nonumber\\&+&|\psi^+\rangle_{ab}\langle 11|+|11\rangle_{ab}\langle \psi^+|)
+ 2 \xi| \psi^+ \rangle_{ab}\langle \psi^+|,\nonumber \\
\label{stateclone}
\end{eqnarray}
where we have used the following notations $\langle Y_0|Y_0\rangle_{x}=\langle Y_1|Y_1\rangle_{x}=\xi$, 
$\langle Y_0 |Q_1\rangle_{x}=\langle Q_0 |Y_1\rangle_{x}=\langle Q_1|Y_0\rangle_{x}=\langle Y_1
|Q_0\rangle_{x}=\frac{\eta}{2}$, $|\psi^+\rangle=\frac{1}{\sqrt{2}}(|01\rangle+|10\rangle)$.
Here, $\eta=1-2\xi$ with $\xi$ being the machine parameter determining the nature of the cloning transformations. 
The output state $\rho_{ab}^{clone}$ is of prime importance as we will investigate the amount of correlations present in it. 
Cloning fidelity is given by the overlap between the real output state $\rho_{b}^{clone}$ 
with the desired output state $|\psi\rangle$. It can be seen that that the cloning fidelity $F_{cl}={\rm Tr}[\rho_{b}^{clone}|
\psi\rangle \langle \psi|]= 1- \xi$ is dependent on the machine parameter $\xi$. 
It has been shown that the BH cloning machine should satisfy the  
inequality $\eta\leq2(\xi-2\xi^{2})^{\frac{1}{2}}$. The relation $\eta=1-2\xi$ reduces the  
inequality $\eta\leq2(\xi-2\xi^{2})^{\frac{1}{2}}$ to the inequality 
$\frac{1}{6}\leq\xi\leq\frac{1}{2}$. Henceforth, we study the different measures of quantum 
correlations in this range of the machine parameter to see how it behaves with the cloning fidelity.
The amount of correlations generated in the process of cloning is given by the difference between the amount of 
correlations in the output modes and the amount of correlations in the same two modes before 
the application of cloning operations. We will denote this difference of correlations as $\Delta^{clone}_{K}= K(\rho^{final}_{ab})-K(\rho^{initial}_{ab})$, for a correlation measure $K(\rho_{ab})$. Here we compute three different correlation 
measures, namely (i) negativity ($N$), (ii) discord ($D$) and (iii)  geometric discord ($DG$) for both the initial input state ($\rho^{initial}_{ab}$) and the final 
output state ($\rho^{final}_{ab}$). Since the Buzek-Hillary cloning machine we start with product states, the respective 
differences $\Delta^{clone}_{N}, \Delta^{clone}_{D}$ and $\Delta^{clone}_{DG}$ of correlations are nothing but the 
amount of correlations  $ N(\rho^{final}_{ab}), D(\rho^{final}_{ab})$ and $DG(\rho^{final}_{ab})$ in the output modes.  
Our first motivation is to see how these different measures of correlation behave with the fidelity of cloning. For this  
purpose, we first express these different measures of the correlation  
$\Delta^{clone}_{N}, \Delta^{clone}_{D}$ and $\Delta^{clone}_{DG}$ in terms of the fidelity  $F_{cl}$ of cloning. 
We rewrite these measures as a function of a variable like the fidelity of cloning $F_{cl}$ and input 
state parameter $\alpha$. 
The expression for $\Delta^{clone}_{N}$ is given by
\begin{eqnarray}
\Delta^{clone}_{N}&=&
\frac{1}{2}\left[2\left\{g_1+\frac{1}{2}g_2f_1\right\}^{\frac{1}{2}}
+\left\{g_1+g_2f_2\right\}^{\frac{1}{2}}
\right.{}\nonumber\\&+&\left.
\left\{g_1+g_2f_3\right\}^{\frac{1}{2}}-1\right],
\end{eqnarray}
where $f_1=|\alpha|^2\beta^2$, $f_2=(1+\frac{1}{2}|\alpha|^2-\alpha^{\ast 2})\beta^2$, 
$f_3=|\alpha|^2(|\alpha|^2+\frac{1}{2}\beta^2)$ (here, $|.|$ denotes absolute value and $\ast$ the 
complex conjugation), $g_1=(F_{cl}-1)^2$ and $g_2=(2F_{cl}-1)^2$. 

\begin{figure}[h]
\begin{center}
\[
\begin{array}{cc}
\includegraphics[height=5.5cm,width=7.5cm]{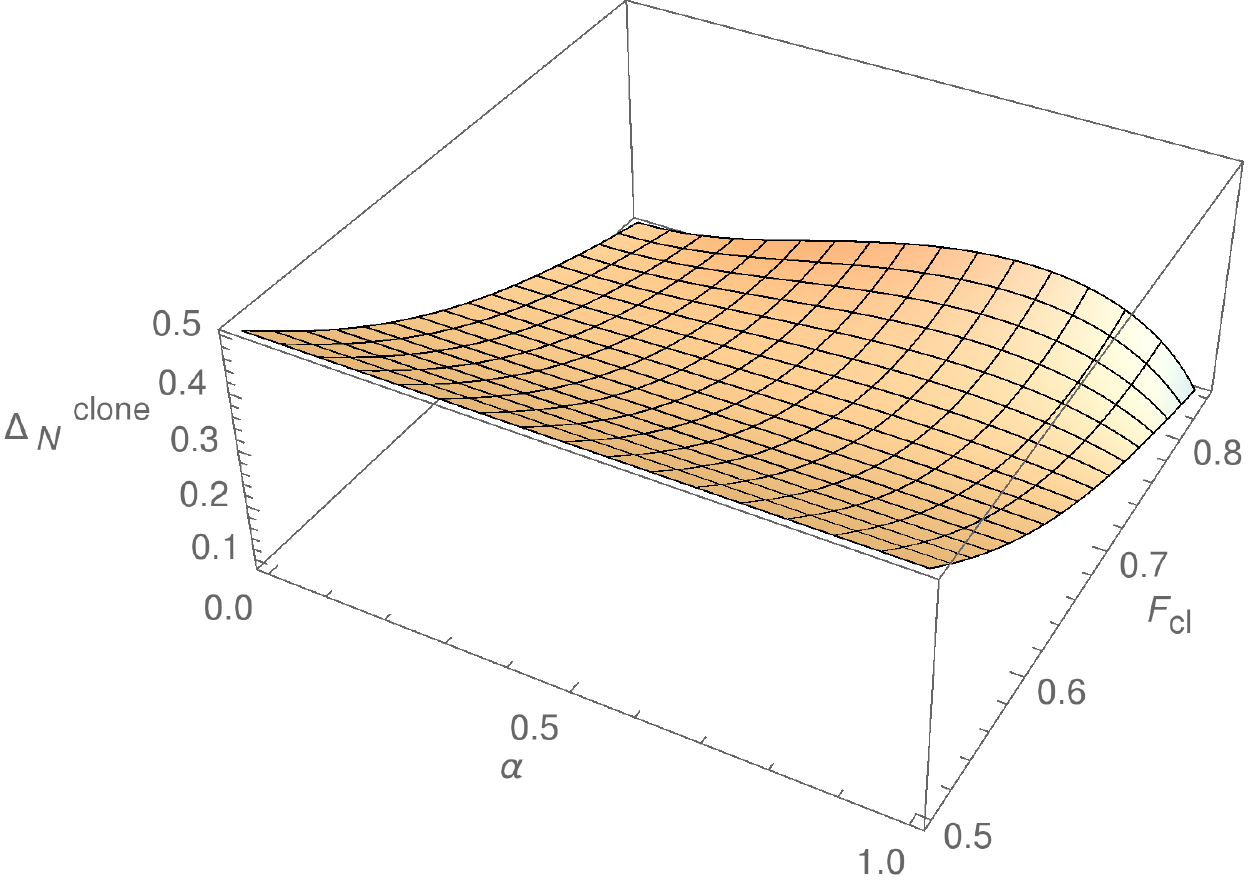}
\end{array}
\]
\caption{(Color online) The plot shows how the correlation measure negativity ($\Delta^{clone}_N$), vary with the input 
parameter $\alpha$ and the fidelity of cloning $F_{cl}$.}\label{neget}
\end{center}
\end{figure}

Similarly the expression for $\Delta^{clone}_{D}$ is given by
\begin{eqnarray}
 \Delta^{clone}_{D}&=&H(F_{cl})+mH(X_{+})-nH(Y_{+}),
%{}\nonumber\\&-&nH(Y_{+},Y_{-}),
\end{eqnarray}
where $H(x)=-x\log_2x-(1-x)\log_2(1-x)$, $X_{+}=\frac{1}{2}(1+\frac{1}{m}\{1+C_+\}^{\frac{1}{2}})$, 
$Y_{+}$ $=\frac{1}{2}(1+ \frac{1}{n}\{4+F_{cl}(7-5F_{cl})+C_-\}^{\frac{1}{2}})$, $C{\pm}=F_{cl}[-2-10\alpha^2
+F_{cl}(1+8\alpha^2)]\pm 3\alpha^2$, $m=n-1$, and $n=\alpha^2+(1-2\alpha^2)F_{cl}$. 

\begin{figure}[h]
\begin{center}
\[
\begin{array}{cc}
\includegraphics[height=5.5cm,width=7.5cm]{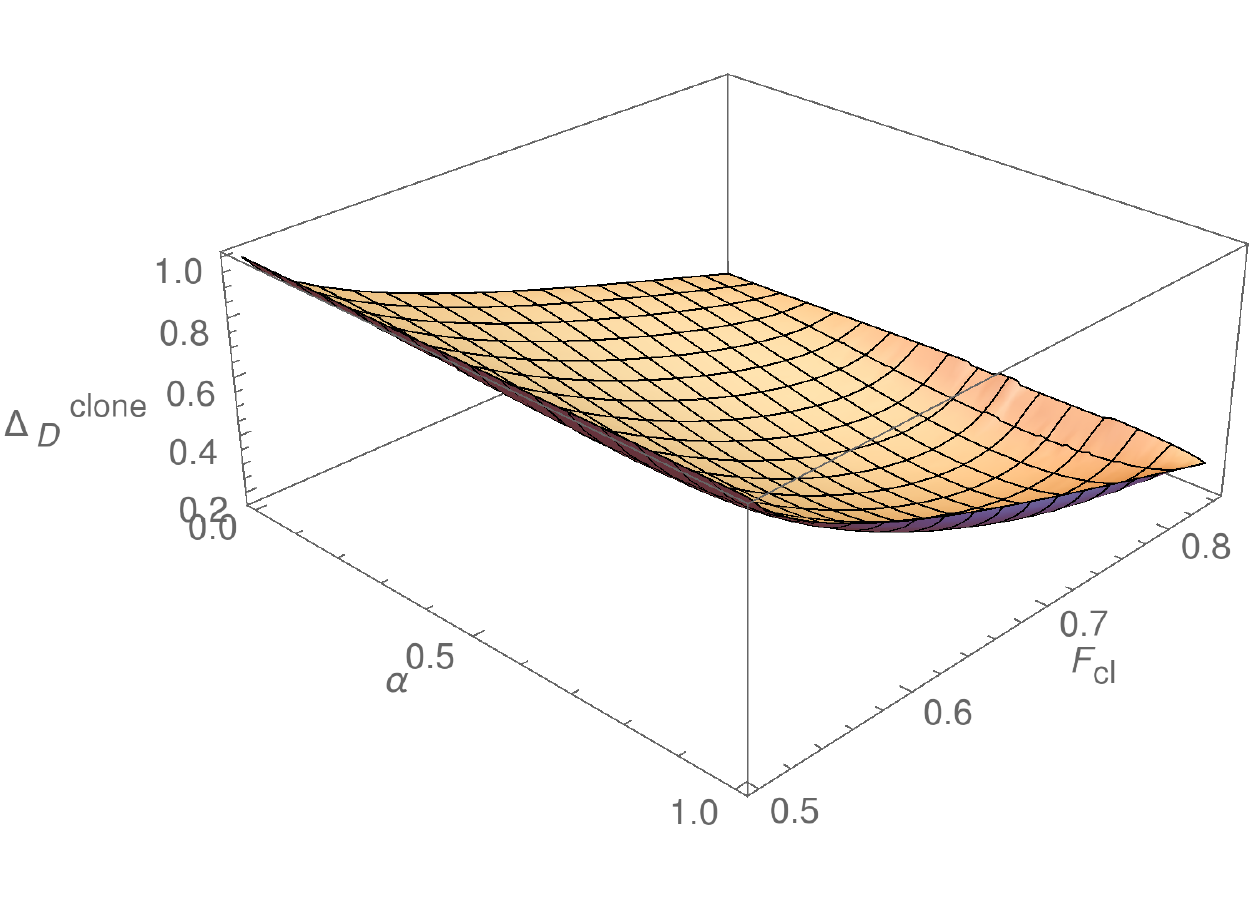}
\end{array}
\]
\caption{(Color online) The plot shows how the correlation measure discord ($\Delta^{clone}_D$) vary with the input 
parameter $\alpha$ and the fidelity of cloning $F_{cl}$.}\label{neget2}
\end{center}
\end{figure}

Lastly, the corresponding expression for the geometric discord is given by
\begin{eqnarray}
 \Delta^{clone}_{DG}=2(\lambda+\lambda_++\lambda_--\max[\lambda,\lambda_+,\lambda_-]),
\end{eqnarray}
where $\lambda=(1-F_{cl})^2$, $\lambda_{\pm}=\frac{1}{2}(3.5-9F_{cl}+6F_{cl}^2\pm 
\sqrt{p-\alpha^2\beta^2q})$,
(here $p=2.25-15F_{cl}+37F_{cl}^2-40F_{cl}^3+16F_{cl}^4$ and $q=5-36F_{cl}+96F_{cl}^2-112F_{cl}^3+48F_{cl}^4$).

\begin{figure}[h]
\begin{center}
\[
\begin{array}{cc}
\includegraphics[height=5.5cm,width=7.5cm]{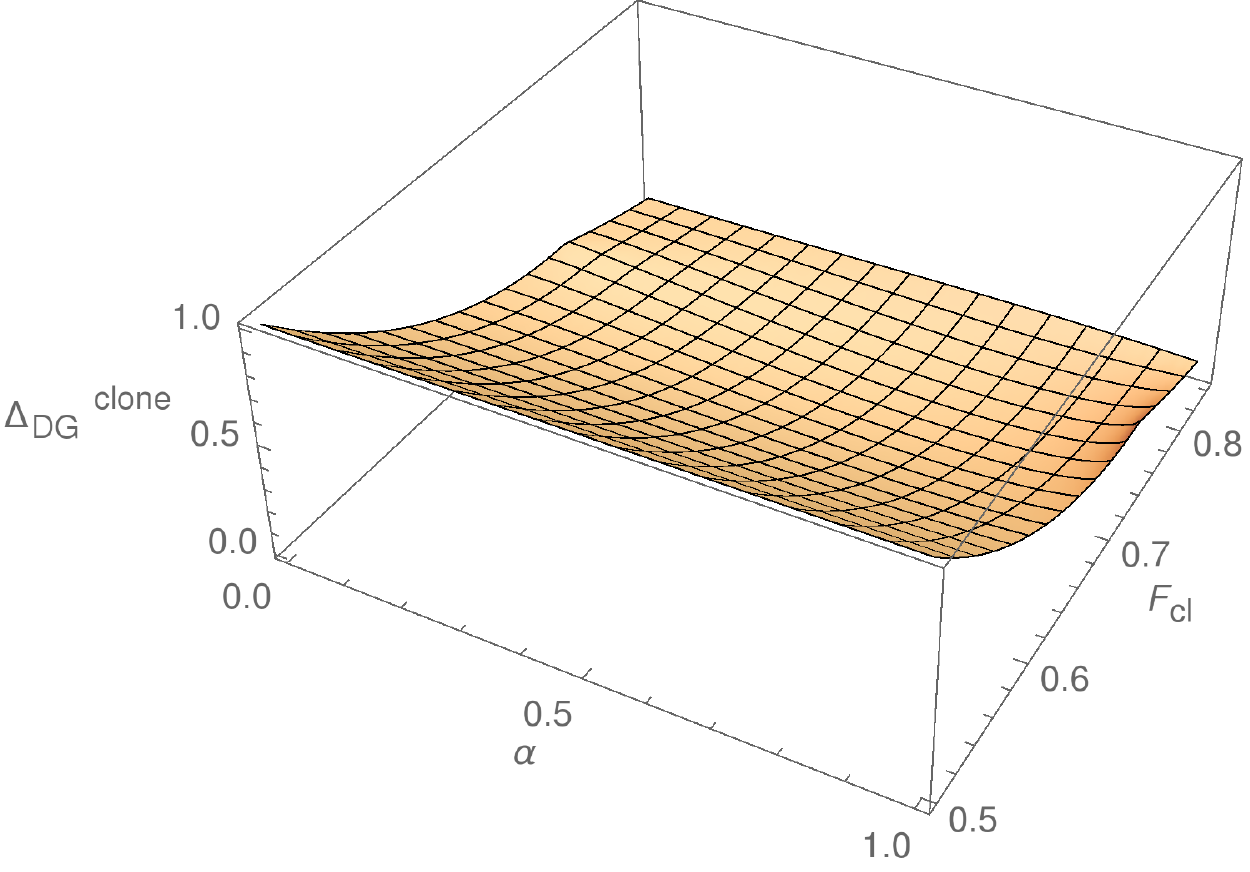}
\end{array}
\]
\caption{(Color online) The plot shows how the correlation measure geometric discord ($\Delta^{clone}_{DG}$) vary with the input 
parameter $\alpha$ and the fidelity of cloning $F_{cl}$.}\label{neget3}
\end{center}
\end{figure}

To have a better insight, we plot these expressions $\Delta^{clone}_{N}, \Delta^{clone}_{D}$ and $\Delta^{clone}_{DG}$ of the correlations 
generated in terms of the fidelity  $F_{cl}$ of cloning and the input state parameter $\alpha$ in
the figures (\ref{neget}, \ref{neget2} \& \ref{neget3}). Since $\xi$ lies in the range
$\frac{1}{6}\leq\xi\leq\frac{1}{2}$, we have the range of the fidelity $\frac{1}{2}\leq F_{cl} \leq\frac{5}{6}$ 
and the range of the input parameter $\alpha$ from $0$ to $1$. 
In figures (\ref{neget}, \ref{neget2} \& \ref{neget3}), we find that the more correlated are states,
the less is the fidelity of cloning. In other words, when we have a 
cloning machine that performs better, the joint output  mode will be poorly correlated.  
Altogether, these plots indicate that the amount of correlations generated in the process 
of cloning plays a vital role in determining the fidelity of cloning. As is evident from  these figures, 
 the more the amount of correlations present in the original and the cloned copy in the output, the more difficult 
it is to copy the information of the original copy in the blank state, because the information gets hidden in the correlations 
between the copies. Though we have considered a particular type of cloning machine to illustrate this phenomenon, we believe
that this phenomenon is independent of the transformation we choose, and is true in general for the process 
of imperfect quantum cloning.
\section{Analysis of the correlations content of the output of a state-dependent deleting machine}
In this section we analyze the correlations generated in the process of quantum deletion which can be thought of as 
the opposite procedure of quantum cloning. As an example, we consider a state-dependent quantum deletion 
machine and study the amount of correlations present 
in the output modes. As in the previous section, we wish to determine the role 
of quantum correlations in regulating the fidelity of deletion. In order to do that, we consider three different 
correlation measures and indeed we see that the physical finding is no different from the cloning.
The action of a state-dependent deleting machine as mentioned in  reference \cite{pati100, Aroy} is given by the 
unitary operation 
\begin{eqnarray}
 \vert \psi  \rangle_{A} \vert \psi  \rangle_{B} \vert A  \rangle_{C} 
 &\rightarrow &\alpha^{2} \vert 0  \rangle_{A} \vert 0  \rangle_{B} \vert A_{0}  \rangle_{C} + \beta^{2} 
 \vert 1  \rangle_{A} \vert 0 \rangle_{B} \vert A_{1}  \rangle_{C} {}\nonumber\\&+&  \alpha \beta( \vert 01 
 \rangle_{AB} + \vert 10  \rangle_{AB})] \vert A \rangle_{C}, \label{deletetr}
\end{eqnarray}
where we start with two copies of the unknown state $|\psi\rangle$ (\ref{state}) with the 
purpose of deleting one copy against the other. Here  $\vert A \rangle_{C}$ is the initial state of the ancilla, 
$\vert A_{0}  \rangle_{C}$ and $\vert A_{1} \rangle_{C}$ are the final states of the ancilla. Moreover, the 
unitarity of the transformation demands the states 
$\vert A \rangle$, $\vert A_{0} \rangle$ and $\vert A_{1} \rangle$ to be orthogonal to each other.
After the application of the deletion transformation given in (\ref{deletetr}) on two copies of $|\psi\rangle$, the output 
reduced density matrix of these two modes takes the form
\begin{eqnarray}
 \rho_{ab}^{del}=|\alpha|^4|00\rangle\langle 00|+|\beta|^4|10\rangle\langle 10|+ 2|\alpha|^2|\beta|^2|\psi^+\rangle\langle \psi^+|,\nonumber \\
 \label{delout}
\end{eqnarray}
where $|\psi^+\rangle=\frac{1}{\sqrt{2}}(|01\rangle+|10\rangle)$. 
The fidelity of the deletion for this machine is given by
$F_{del}=1-|\alpha|^2|\beta|^2$. 
By expressing  the input parameter $|\alpha|^2$ in terms of fidelity $F_{del}$ we have 
$|\alpha|^2=\frac{1}{2}(1\pm\sqrt{4F_{del}-3})$. However the feasible solution for $|\alpha|^2$ is 
$\frac{1}{2}(1-\sqrt{4F_{del}-3})$. Based on the range of $|\alpha|^2$ we find that $F_{del}$ satisfies the relation $\frac{3}{4}\leq F_{del}<1$. This is also consistent with the fact that if we
are given two copies of an unknown qubit, and we perform optimal measurement on both the copies, then we can estimate the 
state with a fidelity $3/4$ \cite{mass} which is also the lower bound of the deletion machine. 
In a similar way, we define the amount of correlations generated in the process of deletion. 
This is given by the difference between the amount of 
correlations in the output modes after the process of deletion, and the amount of correlations in those two modes before 
the application of the deletion operation. We denote this difference of correlations for any correlation measure $K(\rho_{ab})$  
as $\Delta^{del}_{K}=K(\rho^{final}_{ab})-K(\rho^{initial}_{ab})$. We compute various correlation 
measures for both the initial input states and the final 
output states. Since we start with product states having no initial correlation, the amount of correlations generated 
in the process of deletion is the same as the amount of correlations between the output modes. 
We denote these correlations for three different measures (i)  negativity ($N$), (ii)  discord ($D$) 
and (iii) geometric discord ($DG$) by the notations $\Delta^{del}_{N}, \Delta^{del}_{D}$ and $ \Delta^{del}_{DG}$, respectively.
The expression for $\Delta^{del}_{N}$ is given by
\begin{eqnarray}
\Delta^{del}_{N}&=&\frac{1}{2}\left[\frac{(1-a)}{4}\{(1+a)^2+1\}^\frac{1}{2}+(2-a)(1+a)
\right.{}\nonumber\\&-& \left.1\right],
%{}\nonumber\\&&,\right.{}\nonumber\\&+& \left. 
\end{eqnarray}
where $a=\sqrt{4F_{del}-3}$. 

\begin{figure}[h]
\begin{center}
\[
\begin{array}{cc}
\includegraphics[height=5.5cm,width=7.5cm]{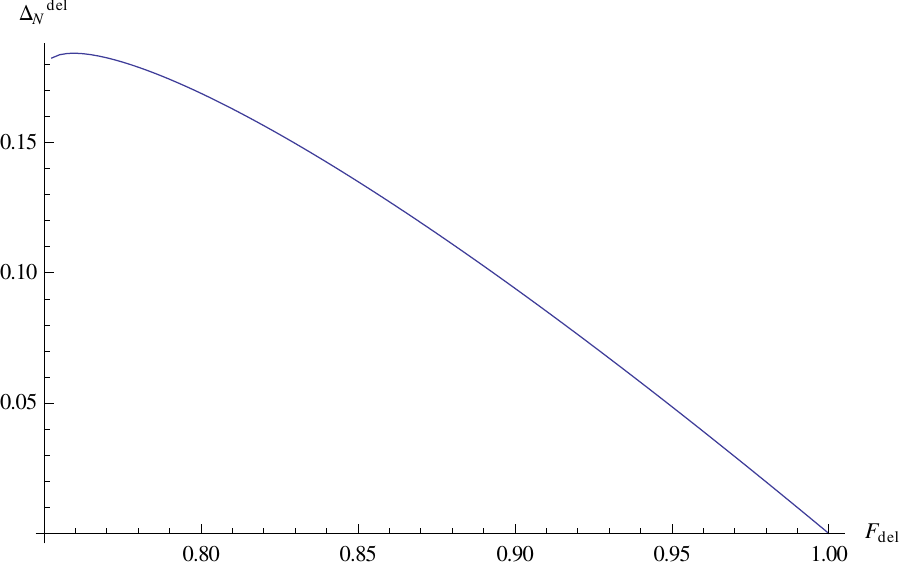}
\end{array}
\]
\caption{(Color online) The figure shows how the correlation measure negativity ($\Delta^{del}_N$) vary with the fidelity of 
deletion ($F_{del}$).}\label{Dinput1}
\end{center}
\end{figure}

Similarly, the expression for $\Delta^{del}_{D}$ is given by
\begin{eqnarray}
 \Delta^{del}_{D}&=&\left(\frac{6}{5}\right)^2\left[H(c)+H(T_+)-h(T_+^2)-h(S_{+},S_{-})\right],\nonumber\\
\end{eqnarray}
where $h(x,y)=-x\log_2x-y\log_2y$, $h(x)=-x\log_2x$, $c=\frac{1}{2F_{del}}(a+1)$, 
$S_{\pm}=\frac{1}{4}(3-2F_{del}+a \pm \{14-2a+4F_{del}(a+5F_{del}-8)\}^\frac{1}{2})$ and $T_{+}=\frac{1}{2}(1-a)$.

\begin{figure}[h]
\begin{center}
\[
\begin{array}{cc}
\includegraphics[height=5.5cm,width=7.5cm]{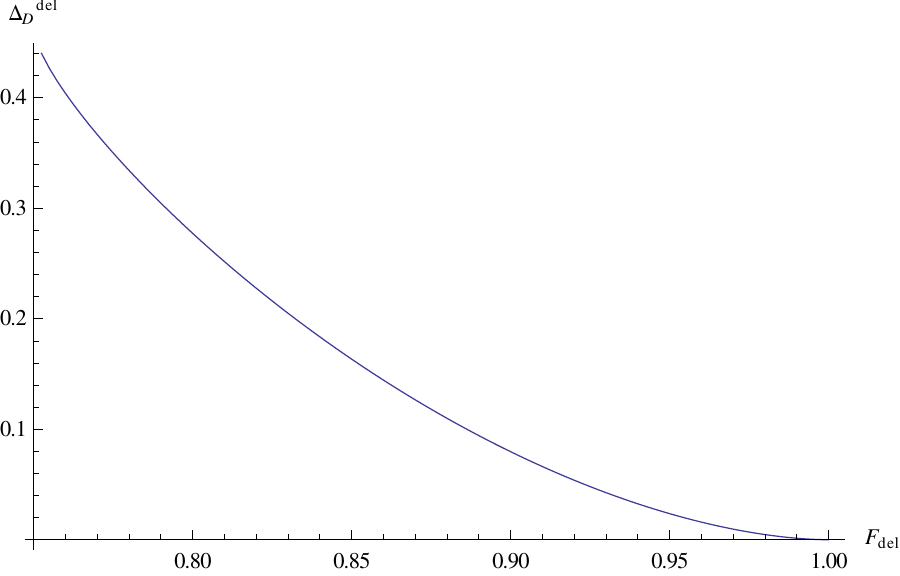}
\end{array}
\]
\caption{(Color online) The figure shows how the correlation measure discord ($\Delta^{del}_D$) vary with the fidelity of 
deletion ($F_{del}$).}\label{Dinput2}
\end{center}
\end{figure}
 
Lastly, the corresponding expression for the geometric discord is given by
\begin{eqnarray}
 \Delta^{del}_{DG}=2(\lambda_0+2\lambda_1-\max[\lambda_0,\lambda_1]),
\end{eqnarray}
where, $\lambda_0=\frac{1}{4}[l_-^2+l_+^2] $, $\lambda_1=K_+^2$, 
$a=\sqrt{4F_{del}-3}$, $l_{\pm}=K_-\pm K_+-1$ and $K_{\pm}=\frac{1}{2}(1-a)\{1\pm \frac{1}{2}(a-1)\}$. 

\begin{figure}[h]
\begin{center}
\[
\begin{array}{cc}
\includegraphics[height=5.5cm,width=7.5cm]{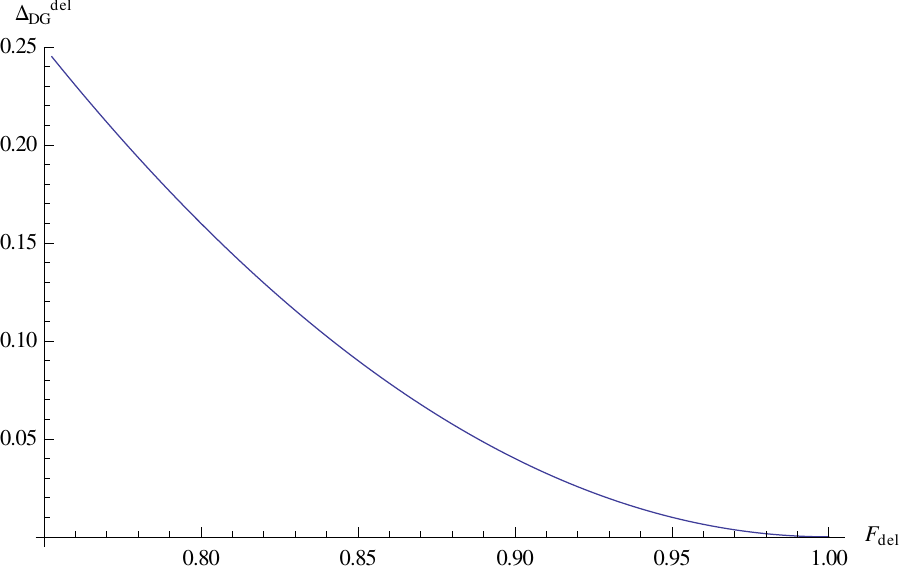}
\end{array}
\]
\caption{(Color online) The figure shows how the correlation measure geometric discord ($\Delta^{del}_{DG}$) vary with the fidelity of 
deletion ($F_{del}$).}\label{Dinput3}
\end{center}
\end{figure}

Now, our aim is to see how  correlations generated in the process controls the fidelity of achieving it. For this, we plot these 
measures with respect to the fidelity of deletion $F_{del}$ in  figures (\ref{Dinput1}, \ref{Dinput2} \& \ref{Dinput3}). 
These figures show that the amount of correlations generated in the process 
varies inversely with the efficiency of carrying out the deletion process successfully. This is very similar to the 
behavior that we have observed in the process of cloning. Our conjecture is that this is  independent of the 
machine we select. This agrees with our physical intuition that the amount of information not available for the 
deletion process is hidden in the correlations between the two modes.
% \begin{widetext} 
% \begin{center}
% \begin{figure}[t]
% \[
% \begin{array}{ccc}
% \includegraphics[height=3.9cm,width=5.5cm]{negativity-delete.pdf}&
% \includegraphics[height=3.9cm,width=5.5cm]{deleting-discord.pdf}&
% \includegraphics[height=3.9cm,width=5.5cm]{geometric-discord-delete.pdf}\\
% (i)&(ii)&(iii)
% \end{array}
% \]
% \caption{The figure shows how the correlation measures ($i$) negativity ($\Delta^{del}_N$), 
% ($ii$)  discord ($\Delta^{del}_D$) and ($iii$) geometric discord ($\Delta^{del}_{DG}$) vary with the fidelity of 
% deletion ($F_{del}$).}
% \label{Dinput1}
% \end{figure}
% \end{center}
% \end{widetext}
\section{Concatenation of Cloning and Deletion -- Correlation Complementarity}
In this section, we consider the successive action of cloning and deletion on an arbitrary quantum state to see that the 
total amount of correlations generated as a result of these two processes is bounded. Here also we find that a
similar thing happens even in the opposite case where cloning is followed by the deletion. These bounds actually show
a new aspect of quantum correlations, i.e., the ``complementarity". We analytically obtain these bounds for 
different measures and exemplify for a particular measure with the help of cloning and deletion machines.
\subsection {Deleting imperfect cloned copies} 
In this subsection, we consider the case where we start with the state to be cloned along with a blank state. The initial state is a product state having no correlation 
at all. After  the cloning operation these two states are no longer uncorrelated and they are given by joint 
density matrix $\rho_{ab}^{final}$. The amount of correlations generated in the process of cloning for a given correlation 
measure $K$ is given by $\Delta^{clone}_{K}=K(\rho^{final}_{ab})-K(|\psi\rangle \otimes |\Sigma\rangle)$. Since the initial 
states are product states, we have $K(|\psi\rangle \otimes |\Sigma\rangle)=0$ and consequently 
$\Delta^{clone}_{K}=K(\rho_{ab}^{final})$. 
$K$ being any correlation measure, is bounded by its maximum and minimum values $K_{max}$ and $K_{min}$ 
respectively.
Now if we delete these imperfect cloned copies in order to get back to its original product form 
$|\psi\rangle \otimes |\Sigma\rangle$, we get a new combined state $\rho_{ab}'$ at the output mode. Then the 
amount of correlations generated in the process is given by   $\Delta^{del}_{K}=K(\rho_{ab}')- K(\rho^{final}_{ab})$ for a 
particular correlation measure $K$. It can be seen that by combining the correlations generated in the cloning 
and deleting process we have
\begin{eqnarray}
\Delta^{clone}_{K}+\Delta^{del}_{K}=K(\rho_{ab}').
\end{eqnarray}
Since the correlation measure $K$ is always bounded by its maximum value $K_{max}$ for any arbitrary state, we have
\begin{eqnarray}
\Delta^{clone}_{K}+\Delta^{del}_{K} \le K_{max}.
\end{eqnarray}
Thus, for the different correlation measures like  negativity ($N$),  discord ($D$) and geometric discord ($DG$) 
we have various bounds for the correlations as given below
\begin{eqnarray}
\Delta^{clone}_{N}+\Delta^{del}_{N} \le \frac{1}{2},\nonumber\\
\Delta^{clone}_{D}+\Delta^{del}_{D} \le 1,\nonumber\\
\Delta^{clone}_{DG}+\Delta^{del}_{DG} \le 1,
\end{eqnarray}
respectively. These bounds together tell us about an intriguing property of quantum correlations which is 
`` complementarity''. The amount of correlations generated in the process of cloning 
is complementary to the amount of correlations generated in the process of deletion. Thus, we can say that when the 
amount of correlations generated in the cloning process is more (less), the amount of correlations for the deletion 
process is less (more).  The above result can be stated differently: it tells us that the better we clone the worse we delete. 
Thus, our conjecture is that this complementarity is not only true for the correlations 
generated but also true for the fidelity of achieving the cloning and deletion process successively. 
\subsubsection{Complementarity for 1$\rightarrow$ 2 cloning, 2 $\rightarrow$ 1 deleting}
Next we exemplify our result with the help of a particular cloning and deleting transformation in the context of a 
specific correlation measure such as the geometric discord ($DG$). We start with an arbitrary quantum state  
$|\psi\rangle$ (\ref{state})  and a blank state 
$|\Sigma\rangle$ initially in the product state.
Then, we apply the universal Buzek-Hillery quantum cloning machine defined by the transformations (\ref{copytr}) 
on $|\psi\rangle$ and on the output of BH copying machine 
we apply the deletion operations defined by 
\begin{eqnarray}
|0\rangle|0\rangle |Q_0\rangle&\rightarrow& |0\rangle|0\rangle|A_0\rangle\nonumber,\\
(|0\rangle|1\rangle+|1\rangle|0\rangle)|Y_i\rangle&\rightarrow& (|0\rangle|1\rangle+|1\rangle|0\rangle)|Y_i\rangle\nonumber,\\
%(|0\rangle|1\rangle+|1\rangle|0\rangle)|Y_1\rangle&\rightarrow& (|0\rangle|1\rangle+|1\rangle|0\rangle)|Y_1\rangle\nonumber,\\
|1\rangle|1\rangle |Q_1\rangle&\rightarrow& |1\rangle|0\rangle|A_1\rangle, \hspace{0.2cm}\mbox{($i=0,1$)}
\end{eqnarray}
to obtain the final output state \cite{adhikari04, Aroy} 
\begin{eqnarray}
 \rho_{ab}'&=&\frac{1}{1+2\xi}(\alpha^2|00\rangle\langle00|+\beta^2|10\rangle\langle 10|+2\xi|\psi^+\rangle\langle\psi^+|)\nonumber\\
%  \nonumber\\&+&2\xi|\psi^+\rangle\langle\psi^+|),
\label{copy123}
\end{eqnarray} 
where, $|\psi^+\rangle=\frac{1}{\sqrt{2}}(|01\rangle+|10\rangle)$ and $\langle A_i|Y_i\rangle=0$. The fidelity of deleting 
imperfect cloned copies is given 
by $F_3=\frac{1+\xi}{1+2\xi}$ \cite{adhikari04} and it ranges from $\frac{3}{4}$ to $\frac{7}{8}$. 
The total correlations generated in the successive process of cloning and deletion is given by the sum of the 
respective correlations
\begin{eqnarray}
\Delta^{T}_{DG} &=& \Delta^{clone}_{DG}+\Delta^{del}_{DG}\nonumber\\
% &= & DG(\rho_{ab}^{clone})-DG(|\psi\rangle \otimes |\Sigma\rangle)\nonumber\\
% &+& DG( \rho_{f})-DG(\rho_{ab}^{clone})
&=&DG(\rho_{ab}').
\end{eqnarray}
The expression for the $\Delta^{T}_{DG}$, i.e., $DG(\rho_{ab}')$ is given by
\begin{eqnarray}
\Delta^{T}_{DG} &=& 2(\lambda_0+2\lambda_1-\max[\lambda_0,\lambda_1]),
\end{eqnarray}
where $\lambda_0=\frac{1}{2}+\sqrt{2}\alpha^4(1-2F_3)^2+2\alpha^2F_3(1-2F_3)-F_3(1-F_3)$ and  
$\lambda_{1}=(1-F_3)^2$.

\begin{figure}[h]
\begin{center}
\[
\begin{array}{cc}
\includegraphics[height=5.5cm,width=7.5cm]{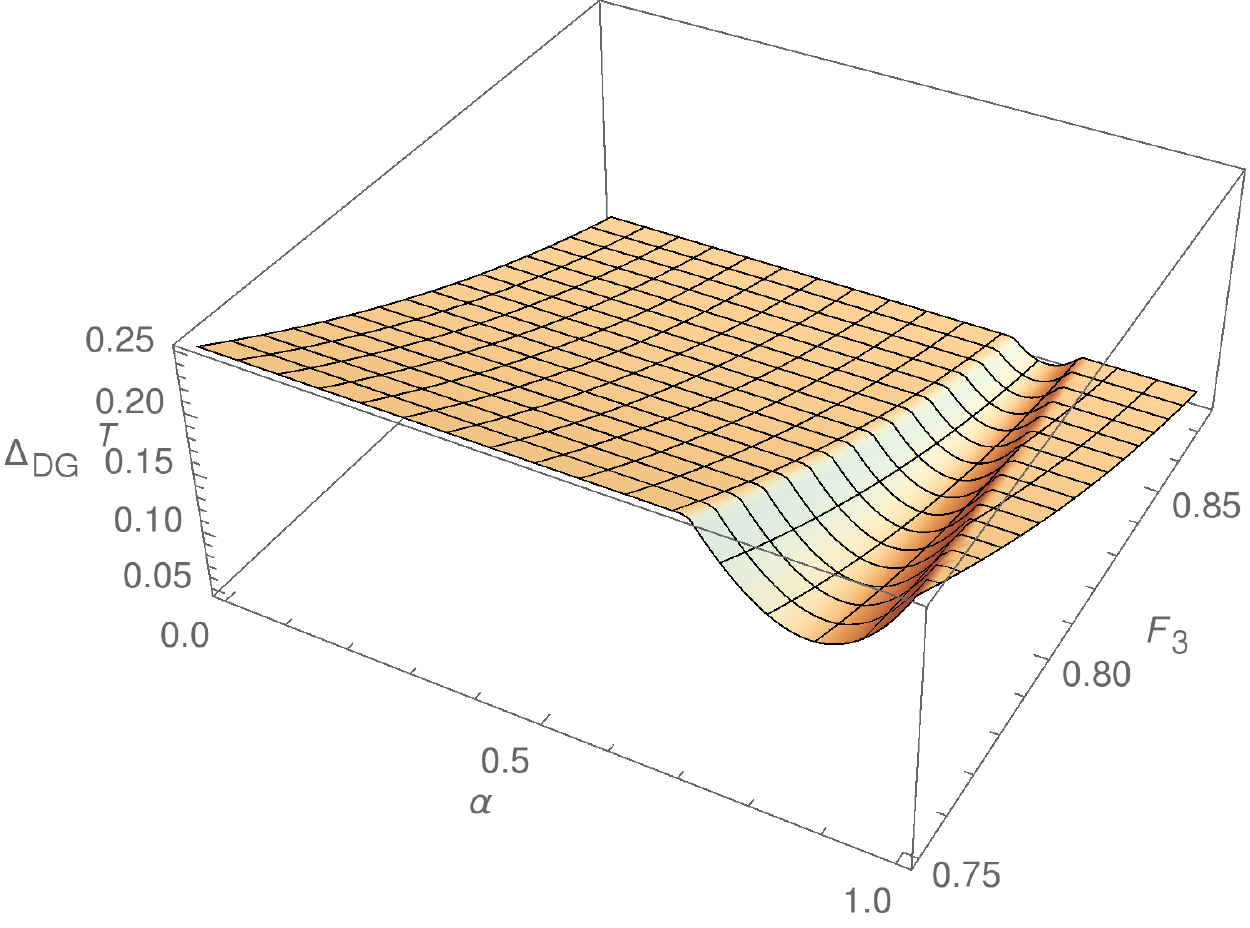}
\end{array}
\]
\caption{(Color online) The figure shows how the total correlations ($\Delta^{T}_{DG}$) for the scheme ``1$\rightarrow$ 2 cloning then 2 $\rightarrow$ 1 deleting'', which 
varies with the input parameter $\alpha$ and the fidelity of deletion $F_3$.}\label{comp1}
\end{center}
\end{figure}
 
Thus, we see that the total correlations generated in the process is given by the correlation content of the 
final state, 
and that it is bounded by its maximum value. Since we adopt  geometric discord as a measure of correlation, the total 
correlations content is bounded by one, i.e., $\Delta^{T}_{DG}<1$. In   figure (\ref{comp1}) we plot the total correlations 
with respect to the machine parameter $\xi$ and the input state parameter $\alpha$ and clearly find that this is always 
bounded by its maximum value one.
% \begin{widetext} 
% \begin{center}
% \begin{figure}[t]
% %\begin{center}
% \[
% \begin{array}{ccc}
% \includegraphics[height=3.9cm,width=5.5cm]{1221dg.jpeg}&
% \includegraphics[height=3.9cm,width=5.5cm]{2112dga.jpeg}&
% \includegraphics[height=3.9cm,width=5.5cm]{2112dg.jpeg}\\
% (i)&(iia)&(iib)
% \end{array}
% \]
% \caption{The figure shows how the total correlations ($\Delta^{T}_{DG}$) for ($i$) the scheme ``1$\rightarrow$ 2 cloning then 2 $\rightarrow$ 1 deleting'', which 
% varies with the input parameter $\alpha$ and the fidelity of deletion $F_3$, and for   
% ($ii$) the scheme ``2$\rightarrow$ 1 deleting then 1 $\rightarrow$ 2 cloning'', ($iia$) $\Delta^{T}_{DG}$ of equation (\ref{123u})
% and ($iib$) $\Delta^{T}_{DG}$ of equation (\ref{123u2})  varies with input parameter $\alpha$ and the cloning machine parameter $\xi$.}
% \label{comp1}
% \end{figure}
% \end{center}
% \end{widetext}
\subsubsection{Complementarity for 1$\rightarrow$ N cloning, N $\rightarrow$ M deleting}
Next we extend our result to a more general situation, where we first create $N$ copies from a single 
copy with the help of a ``$1\mapsto N$''-cloning machine. Then we use a ``$N\mapsto M$ ($N>M$) ''deleting machine to produce 
$M$ distorted copies of the input state at the output port. 
%process only as a particular case of ``$M'\mapsto N$-Clone then $N\mapsto M$ Deleting ($M,M'< N$)'' because it 
%will complete our purpose. 
First of all, we apply ``$1\mapsto N$'' Cloning machine on an arbitary input state $\ket{\psi}$ 
(\ref{state}). 
We use the result of Gisin and Massar who first generalized the Buzek-Hillery's $1\mapsto 2$ cloning machine to $M'\mapsto N$ ($M'<N$)
\cite{mas}. 
%They considered the input state of the quantum cloning machine (QCM) as $\ket{\psi}=\alpha\ket{0}+\beta\ket{1}$. 
Now for $M'=1$, the unitary operator ($U_{1,N}$) for $1\mapsto N$ cloning machine is given by
\begin{eqnarray}
U_{1,N}\ket{0}\otimes R&=&\displaystyle\sum_{j=0}^{N-1}\alpha_j\ket{(N-j)0,j1}\otimes R_j,\nonumber\\
U_{1,N}\ket{1}\otimes R&=&\displaystyle\sum_{j=0}^{N-1}\alpha_{N-1-j}\ket{(N-1-j)0,(j+1)1}\otimes R_j,\nonumber\\
\end{eqnarray}
where $R$ denotes initial combined state of the copying machine and $(N-1)$ blank copies. Here $R_j$ are orthonormalized internal 
states of the quantum cloning machine. Here, $\alpha_j=\sqrt{2(N-j)/N(N+1)}$ and we have denoted $\ket{(N-j)0,j1}$ as the symmetric and 
normalized state.
After ``$1\mapsto N$''-cloning operation is over, we use the ouput of cloning machine as an input 
to a ``$N\mapsto M$'' deleting machine . The action of the deleting machine is given by the transformations \cite{pati100,Aroy},
\begin{eqnarray}
\ket{0}^{\otimes N}\ket{R_0}&\mapsto  &\ket{0}^{\otimes N}\ket{A_0},\nonumber\\
 \ket{(N-j)0,j1}\ket{R_j}&\mapsto  &\ket{(N-j)0,j1}\ket{R_j},\hspace{0.2cm}\mbox{$j\neq 0$}\nonumber\\
% \ket{(N-1-j)0,(j+1)1}\ket{R_j}&\mapsto  &\ket{(N-1-j)0,(j+1)1}\ket{R_j},\hspace{0.2cm}\mbox{$j\neq N-1$}.\nonumber\\
\ket{1}^{\otimes N}\ket{R_{N-1}}&\mapsto  &\ket{1}^{\otimes M}\ket{0}^{\otimes (N-M)}\ket{A_1},
\label{delnm}
\end{eqnarray}
where $\ket{A_0}$, $\ket{A_1}$ are machine states at the output port of the deleting machine.
Combining these two machine, the complete transformation of $\ket{\psi}$ is given by
\begin{eqnarray}
\ket{\psi}&\mapsto & \alpha \left[\alpha_0\ket{N0}\otimes A_0+\displaystyle\sum_{j=1}^{N-1}\alpha_j\ket{(N-j)0,j1}\otimes R_j\right]\nonumber\\ &+&\beta\left[\displaystyle\sum_{j=0}^{N-2}\alpha_{N-1-j}\ket{(N-1-j)0,(j+1)1}\otimes R_j \right.\nonumber\\ &+& \left.\alpha_0\ket{M1(N-M)0}\otimes A_1 \right],
\end{eqnarray}
where $\bra{A_i}R_j\rangle=0$, $\bra{A_i}A_j\rangle=\delta_{ij}$, $\bra{R_i}R_j\rangle=\delta_{ij}$ 
 ($\delta_{ij}$ is  Kronecker delta). The density matrix at the output port after tracing out the machine states is given by
\begin{widetext}
\begin{eqnarray}
\rho &=& \alpha^2 \left[\alpha_0^2\ket{N0}\bra{N0}+
\displaystyle\sum_{j=1}^{N-1}\alpha_j^2\ket{(N-j)0,j1}\bra{(N-j)0,j1}\right]\nonumber\\&+&\displaystyle\sum_{i=1}^{N-1}\displaystyle\sum_{j=0}^{N-2}\alpha_i\alpha_{N-1-j}(\alpha\beta^*\ket{(N-i)0,i1}\bra{(N-1-j)0,(j+1)1}
%\nonumber\\&+&
+\alpha^*\beta\ket{(N-1-j)0,(j+1)1}\bra{(N-i)0,i1})\delta_{ij},\nonumber\\ &+&\beta^2\left[\displaystyle\sum_{j=0}^{N-2}\alpha_{N-1-j}^2\ket{(N-1-j)0,(j+1)1}\bra{(N-1-j)0,(j+1)1} 
% \right.\nonumber\\ &+& \left.
+\alpha_0^2\ket{M1(N-M)0}\bra{M1(N-M)0} \right].
\end{eqnarray}
\end{widetext}
After tracing out rest of modes, the reduced density matrix of the first mode is is given by
\begin{eqnarray}
 \rho^a= \displaystyle\sum_{i=0}^{N-1}\left\{\frac{(N-i)}{N} C(N,N-i)\alpha^2 +\frac{i}{N}C(N,i)\beta^2\right\}\alpha_i^2\ket{0}\bra{0}
\nonumber\\ +\displaystyle\sum_{i=0}^{N-1}\left\{\frac{i}{N}C(N,i)\alpha^2+\frac{(N-i)}{N}C(N,N-i)\beta^2\right\}\alpha_i^2\ket{1}\bra{1},\nonumber
\end{eqnarray}
where $C(x,y)=\frac{x!}{y! (x-y)!}$.

\begin{figure}[h]
\begin{center}
\[
\begin{array}{cc}
\includegraphics[height=5.5cm,width=7.5cm]{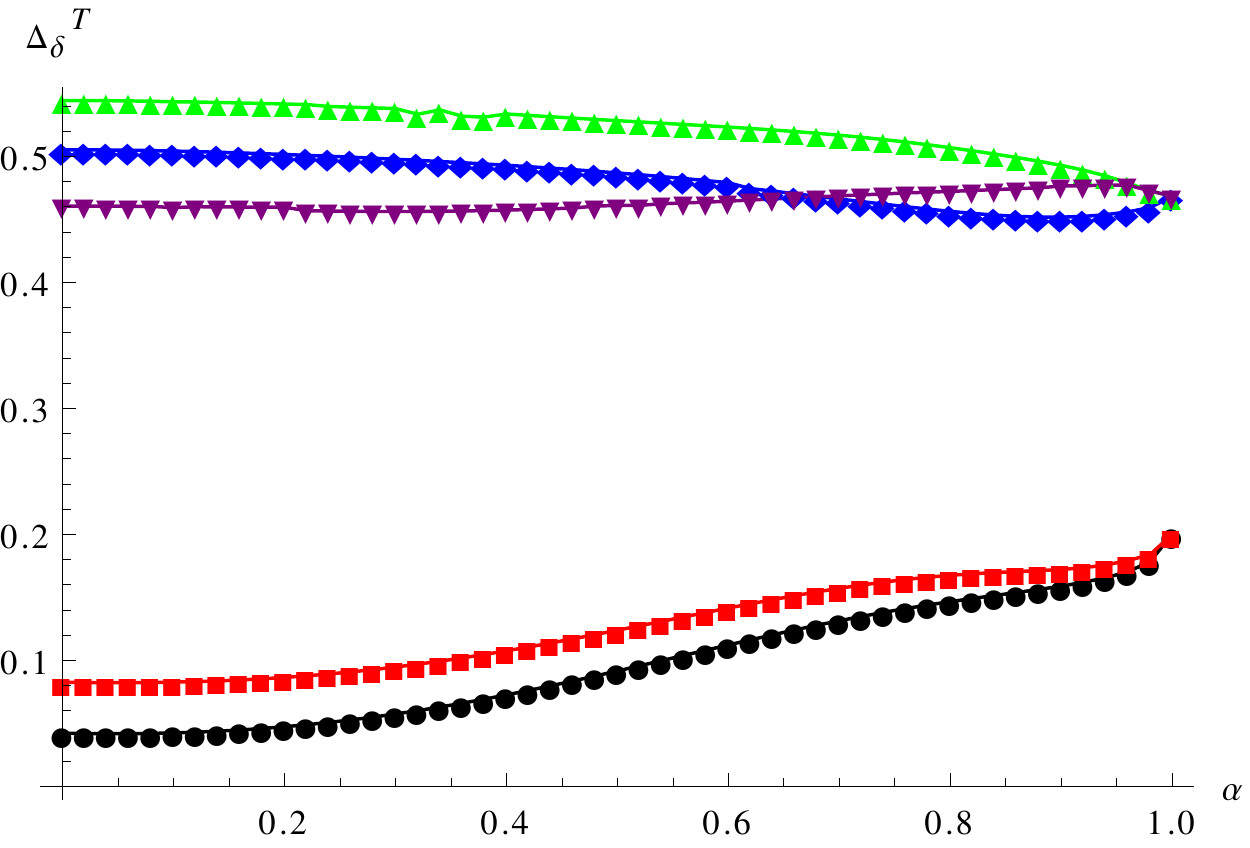}
\end{array}
\]
\caption{(Color online) Quantum correlations $\Delta_{\delta}^{T}$ versus input state parameter $\alpha$ for the processes (a) $1\mapsto3$ cloning then $3\mapsto 1$ deleting ($\bullet$), (b) $1\mapsto3$ cloning then $3\mapsto 2$ deleting ({\color{red} $\blacksquare$}), (c) $1\mapsto4$ cloning then $4\mapsto 1$ deleting ({\color{blue} $\blacklozenge$}), (d) $1\mapsto4$ cloning then $4\mapsto 2$ deleting ({\color{green} $\blacktriangle$}) and (e) $1\mapsto4$ cloning then $4\mapsto 3$ deleting ({\color{purple(html/css)} $\blacktriangledown$}).}
\label{figrty}
\end{center}
\end{figure}

Since, we know that for a multiqubit state, there is no unique way to quantify quantum correlations present in the state. 
For that reason we have taken a simple approach and 
have considered bipartite discord as a measure of quantum correlations. 
The bipartite discord of the $N$-qubit state $\rho_{1,...,N}$ (for the partition $(i,\bar{i})$) is defined as,
\begin{eqnarray}
  D(i|\bar{i})= \min_{\Pi_{\bar{i}_{j}}}\{S(\rho_{\bar{i}})+S(\rho_{i|\bar{i}})-S(\rho_{1,...,N})\},\label{discorn}
\end{eqnarray}
where $S(\rho_{i|\bar{i}})=\sum_j p_j S(\rho_{i|j})$ is the average
of the entropies of 
states $\rho_{i|j}=\frac{1}{p_j}\Tr_{\bar{i}}[(\mathbb{I}_i\otimes\Pi_{\bar{i}_{j}})\rho_{1,...,N}(\mathbb{I}_i\otimes\Pi_{\bar{i}_{j}})]$ 
with corresponding probability $p_j=\Tr[(\mathbb{I}_i\otimes\Pi_{\bar{i}_{j}})\rho_{1,...,N}(\mathbb{I}_i\otimes\Pi_{\bar{i}_{j}})]$. 
Here  $\Pi_{\bar{i}_{j}}$'s are all possible $(N-1)$ qubits projective measurement operators. The bipartite discord given in equation 
(\ref{discorn}) is to be minimized over all possible projective measurements $\Pi_{\bar{i}_{j}}$. In this paper we have performed 
projective measurements upto three qubits.  For single qubit measurement we have used the measurement bases given in 
\cite{oli,hen}. For two and three qubit measurements we have followed the reference \cite{ind1}.  
The bipartite discord (see equation (\ref{discorn})) is not symmetric under the exchange of qubit. 
So, the average correlations present in the $N$-qubit state is given by,
\begin{equation}
 \delta(\rho_{1,...,N})=\frac{1}{N}\displaystyle\sum_{i=1}^N D(i|\bar{i}).
\end{equation}
So the total correlations generated in this process is $\Delta_{\delta}^T=\Delta_{\delta}^{clone}+\Delta_{\delta}^{del}=\delta(\rho)$. 
In figure (\ref{figrty}), we have plotted the total correlations $\Delta_{\delta}^T$ generated during cloning and deleting against the input state 
parameter ($\alpha$) to show the complementary nature of correlations production in this two processes, i.e., 
$\Delta_{\delta}^{clone}+\Delta_{\delta}^{del}\leq 1$. In the figure, we have  looked for the 
correlations generated in each of the dual processes: 
(a) $1\mapsto3$ cloning then $3\mapsto 1$ deleting, (b) $1\mapsto3$ cloning then $3\mapsto 2$ deleting,
(c) $1\mapsto4$ cloning then $4\mapsto 1$ deleting, (d) $1\mapsto4$ cloning then $4\mapsto 2$ deleting 
and (e) $1\mapsto4$ cloning then $4\mapsto 3$ deleting. It is evident from the figure itself, that the total 
correlations generated as a consequence of dual processes in each of these cases is bounded. Thus even in a 
most general setting of multiple qubits the correlations generated in each of cloning and deleting process are 
complementary in nature.   
% \begin{widetext} 
% \begin{center}
% \begin{figure}[t]
% %\begin{center}
% \[
% \begin{array}{ccc}
% \includegraphics[height=3.5cm,width=5cm]{clone2delete-gdiscord.eps}&
% \includegraphics[height=3.5cm,width=5cm]{delete2clone-gdiscordaa.eps}&
% \includegraphics[height=3.5cm,width=5cm]{delete2clone-gdiscordbb.eps}\\
% (i)&(iia)&(iib)
% \end{array}
% \]
% \caption{The figure shows how the total correlation ($\Delta^{T}_{DG}$) for ($i$) the scheme ``deleting imperfect cloned
% copies'', which 
% varies with the input parameter $\alpha$ and the fidelity of deletion $F_3$, and for   
% ($ii$) the scheme ``cloning of imperfect deleted copies'', ($iia$) $\Delta^{T}_{DG}$ of equation (\ref{123u})
% and ($iib$) $\Delta^{T}_{DG}$ of equation (\ref{123u2})  varies with input parameter $\alpha$ and the cloning machine parameter $\xi$.}
% \label{comp1}
% \end{figure}
% \end{center}
% \end{widetext}
\subsection{Cloning of imperfect deleted copies} 
In this subsection we carry out the reverse process where we perform deleting first and then clone the imperfect deleted 
copies. We start with two identical copies of an unknown quantum state $|\psi\rangle$ 
(\ref{state}). Initially, there is no correlation between these two states as 
they are in the product form. Consequently, we can write the correlation content of these states for a given correlation 
measure $K$ as $K(|\psi\rangle \otimes |\psi\rangle)=0$. However, after the deletion operation they are no longer uncorrelated.
Instead, we obtain a correlated two qubit state $\rho_{ab}^{del}$. The amount of  correlations 
generated in the process of deletion is given by the difference of the correlations of the final and the initial 
states, i.e., $\Delta^{del}_{K}=K(\rho_{ab}^{del})-K(|\psi\rangle \otimes |\psi\rangle)=K(\rho_{ab}^{del})$. Next, we apply the cloning 
transformations on the combined state $\rho_{ab}^{del}$ in order to get back to the initial identical copies of the state 
$|\psi\rangle $. However, due to the imperfectness of the process we get a mixed state $\rho^{del}_{clone}$ 
at the output port. 
The amount of correlations generated in the process is given by the difference of the correlations of the states 
$\rho_{ab}^{del}$ and $\rho^{del}_{clone}$, i.e., $\Delta^{clone}_{K}=K(\rho^{del}_{clone})-K(\rho_{ab}^{del})$. The total  correlations generated in the process of cloning and deletion is given by 
\begin{eqnarray}
\Delta^{del}_{K}+\Delta^{clone}_{K}=K(\rho^{del}_{clone}).
\end{eqnarray}
Since for a given correlation measure $K$ the correlations of a particular state is always bounded by its maximum and 
minimum value $K_{max}$ and $K_{min}$, we will get back the same bound on the total correlations generated, i.e.,
\begin{eqnarray}
\Delta^{del}_{K}+\Delta^{clone}_{K} \le  K_{max},
\end{eqnarray}
irrespective of whether we delete and then clone or we clone first and then delete. This once again 
establishes the same complementarity in terms of the correlations generated in the process of cloning and deletion. 
The complementarity of quantum correlations are 
independent of whether we apply cloning or deletion first.
\subsubsection{Complementarity for 2$\rightarrow$ 1 deleting, 1 $\rightarrow$ 2 cloning}
 
Next, we give an example of the complementarity phenomenon in this case with the help of a particular deleting 
and cloning machine in the context of a specific correlation measure, namely geometric discord ($DG$). 
Here we start with two identical copies of the state $|\psi\rangle$ and we apply the  
quantum deletion machine defined in equation (\ref{deletetr}) which results in a two qubit state $\rho_{ab}^{del}$  
(see equation (\ref{delout})). Then, we apply BH cloning operation on the state $\rho_{ab}^{del}$ which will give us 
two output states as
$\rho_{aa^{\prime}}={\rm Tr}_{b}[(U_{BH}\otimes I)(\rho_{ab}^{del}|0\rangle_{a^{\prime}}\langle 0|)(U_{BH}\otimes I)^{\dagger}]$ and 
$\rho_{bb^{\prime}}={\rm Tr}_a[(I\otimes U_{BH})(\rho_{ab}^{del}|0\rangle_{b^{\prime}}\langle 0|)(I\otimes U_{BH})^{\dagger}]$. 
The density operators $\rho_{aa^{\prime}}$ and $\rho_{bb^{\prime}}$ are given by
\begin{eqnarray}
 \rho_{aa^{\prime}}&=&(1-2\xi)(\alpha^2|00\rangle\langle00|+\beta^2|11\rangle\langle 11|)
 \nonumber\\&+& 2\xi|\psi^+\rangle\langle\psi^+|, \hspace{0.2 cm}\mbox{and}\nonumber\\
 \rho_{bb^{\prime}}&=&(1-2\xi)\{(1-\alpha^2\beta^2)|00\rangle\langle00|+\alpha^2\beta^2|11\rangle\langle 11|\}
 \nonumber\\&+& 2\xi|\psi^+\rangle\langle\psi^+|.
 \label{del2col}
\end{eqnarray}
The total correlations generated in the successive process of deletion and cloning is given by the sum of the 
respective correlations. Here, we obtain the total correlations in terms of the measure geometric discord ($DG$) as
\begin{eqnarray}
\Delta^{T}_{DG}&=&\Delta^{del}_{DG}+\Delta^{clone}_{DG}\nonumber\\
&=&DG(\rho^{del}_{clone}).
\end{eqnarray}
In this case $\rho^{del}_{clone}$ are $\{\rho_{aa^{\prime}},\rho_{bb^{\prime}}\}$. Hence, the total correlations for the state 
$\rho_{aa^{\prime}}$ is given by
\begin{equation}
 \Delta^{T}_{DG}=2(\lambda_0+2\lambda_1-\max[\lambda_0,\lambda_1]),\label{123u}
\end{equation}
where $\lambda_0=\frac{1}{4}[L^2+(L-2\xi)^2$], $\lambda_{1}=\xi^2$ and $L=(1-2\xi)(\alpha^2-\beta^2)$. 

\begin{figure}[h]
\begin{center}
\[
\begin{array}{cc}
\includegraphics[height=5.5cm,width=7.5cm]{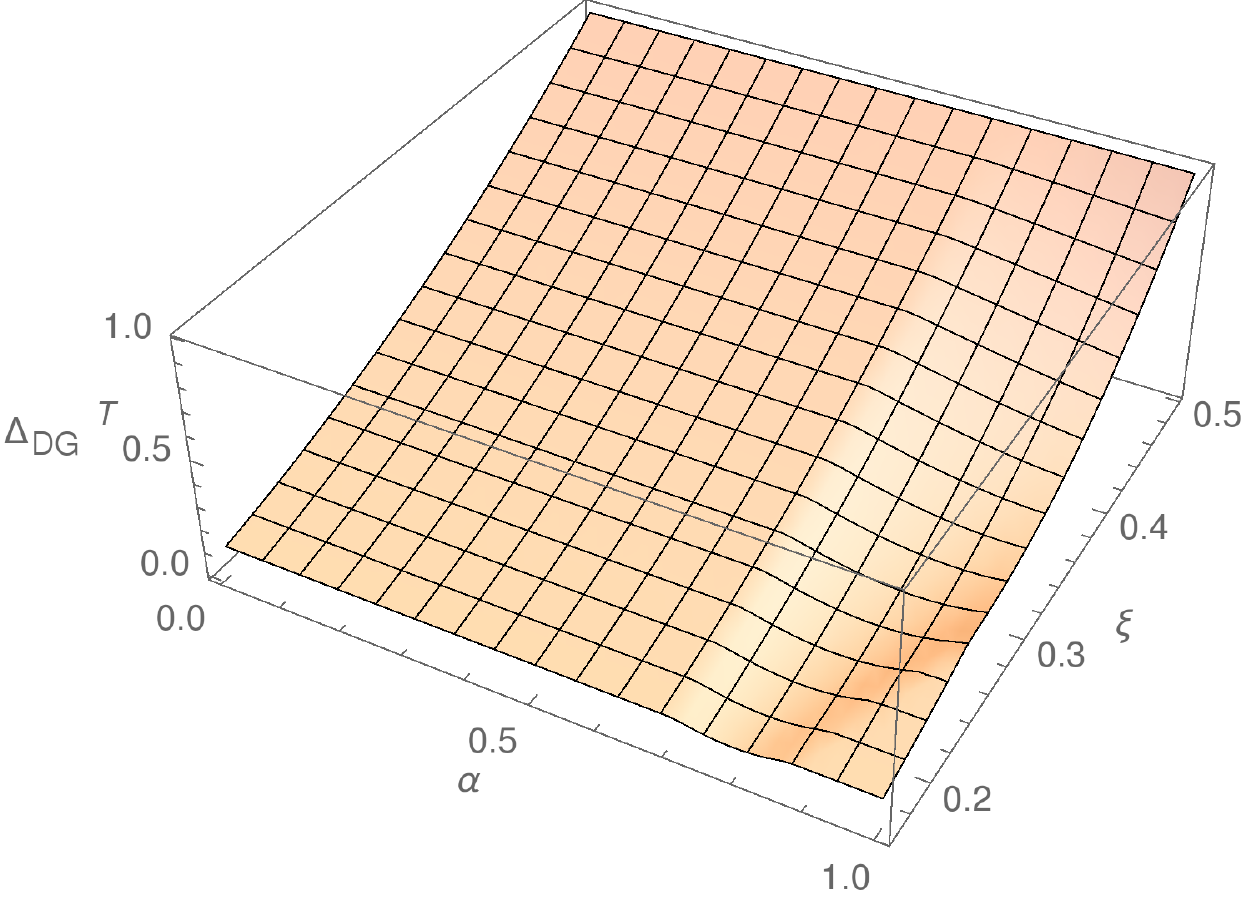}
\end{array}
\]
\caption{(Color online) The figure shows how the total correlations ($\Delta^{T}_{DG}$) of equation (\ref{123u}) for the scheme ``2$\rightarrow$ 1 deleting then 1 $\rightarrow$ 2 cloning'' varies with input parameter $\alpha$ and the cloning machine parameter $\xi$.}\label{comp2}
\end{center}
\end{figure}

Similarly, for 
$\rho_{bb^{\prime}}$ we find
\begin{equation}
 \Delta^{T}_{DG}=2(\lambda_0+2\lambda_1-\max[\lambda_0,\lambda_1]),\label{123u2}
\end{equation}
where $\lambda_0=\frac{1}{4}[J^2+(1-4\xi)^2$], $\lambda_{1}=\xi^2$ and $J=(1-2\xi)(1-2\alpha^2\beta^2)$. 

\begin{figure}[h]
\begin{center}
\[
\begin{array}{cc}
\includegraphics[height=5.5cm,width=7.5cm]{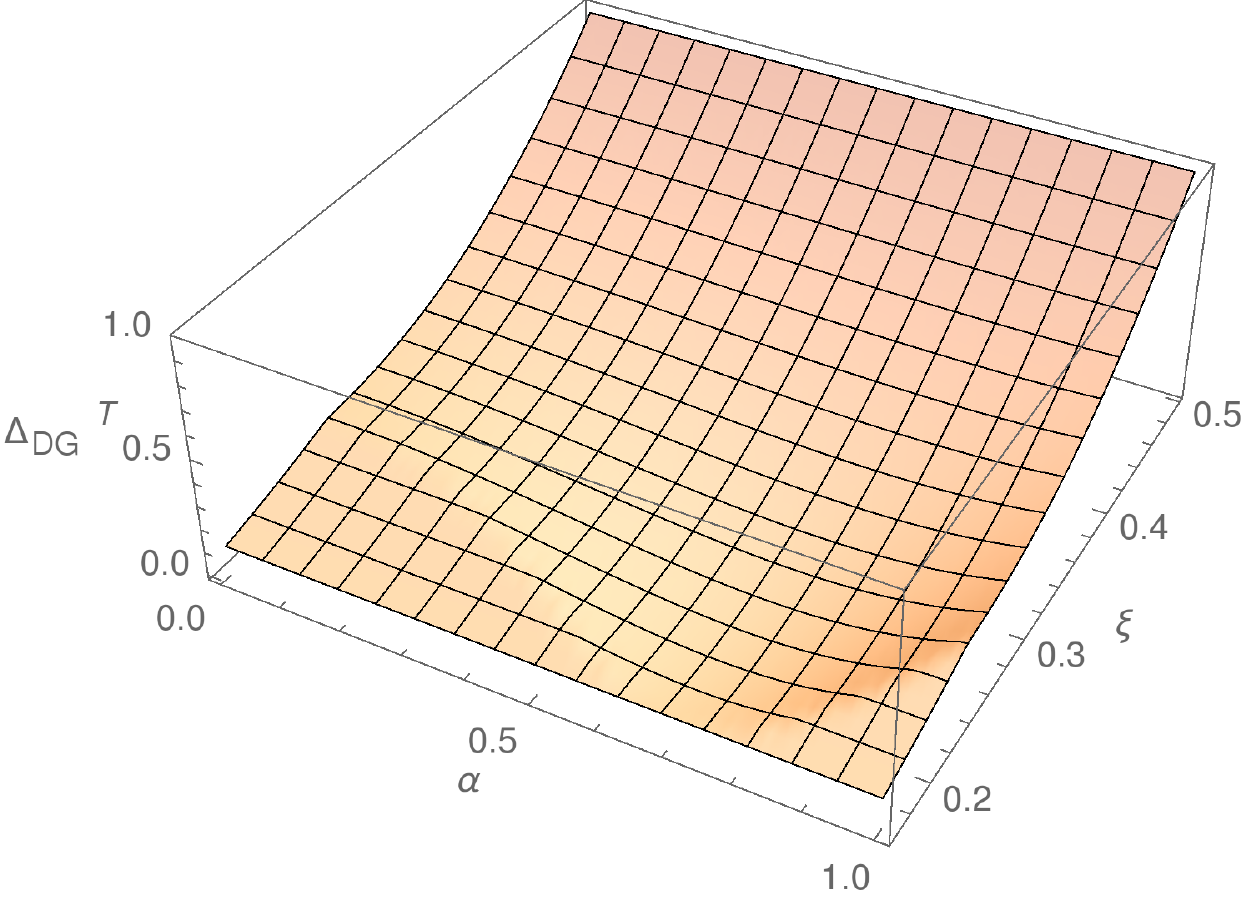}
\end{array}
\]
\caption{(Color online) The figure shows how the total correlations ($\Delta^{T}_{DG}$) of equation (\ref{123u2}) for the scheme ``2$\rightarrow$ 1 deleting then 1 $\rightarrow$ 2 cloning'' varies with input parameter $\alpha$ and the cloning machine parameter $\xi$.}\label{comp3}
\end{center}
\end{figure}
 
As in the previous process, here too the total correlations are given by the correlations of the final 
state.
In figures (\ref{comp2} \& \ref{comp3}), we plot the total correlations $\Delta^{T}_{DG}$ against the input parameter $\alpha$ to find that this is always 
bounded by its maximum value one, i.e., $\Delta^{T}_{DG}\leq 1$.\\
\subsubsection{Complementarity for N$\rightarrow$ 1 deleting, 1 $\rightarrow$ M cloning}
\begin{figure}[h]
\begin{center}

\[
\begin{array}{cc}
\includegraphics[height=5.5cm,width=7.5cm]{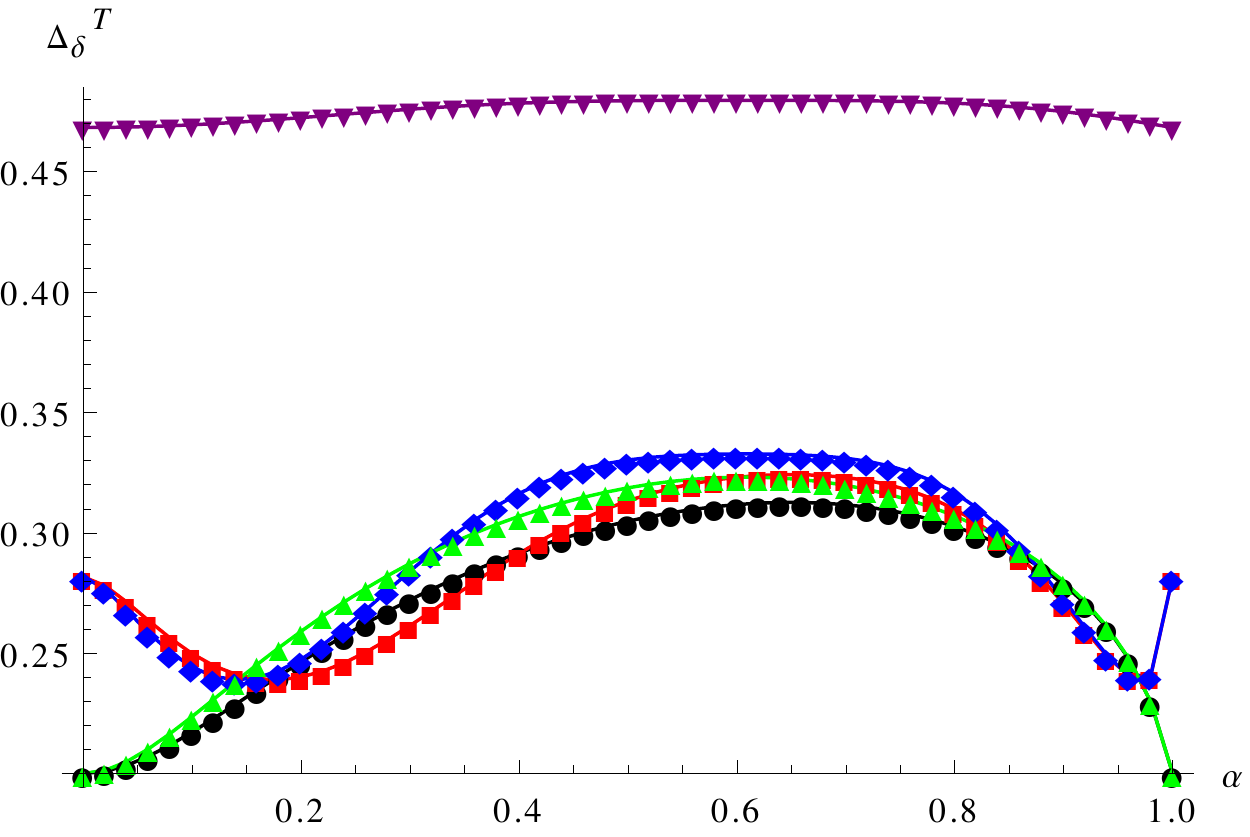}
\end{array}
\]
\caption{(Color online) Quantum correlations $\Delta_{\delta}^{T}$ varses input state parameter $\alpha$ for the processes (a) $3\mapsto1$ deleting then $1\mapsto 2$ cloning ({\color{red} $\blacksquare$}), (b) $3\mapsto1$ deleting then $1\mapsto 3$ cloning ($\bullet$) , (c) $4\mapsto1$ deleting then $1\mapsto 2$ cloning ({\color{blue} $\blacklozenge$}), (d) $4\mapsto1$ deleting then $1\mapsto 3$ cloning ({\color{green} $\blacktriangle$}) and (e) $4\mapsto1$ deleting then $1\mapsto 4$ cloning ({\color{purple(html/css)} $\blacktriangledown$}).}
\label{1nnm1}
\end{center}
\end{figure}

Further we move on to much more general setting where we start with the application of ``$N\mapsto 1$''
deleting machine on $N$ copies of the state $\ket{\psi}$ 
(\ref{state}) to produce 
a distorted state at the output port. Let say, at the output port we will have the state $\rho_{a_1,..,a_N}^{del}$ after tracing out the machine states, where $a_1$ is the 'undeleted mode' and $a_2,..,a_N$ are the 'deleted modes'. 
In the next step, we take the state ($\rho_{a_i}^{del}$; $a_i\neq a_1$) of $\rho_{a_1,..,a_N}^{del}$ 
as an input to  ``$1\mapsto M$'' cloning process. 
% It is to be noted that here in this case we are considering only the modes $a_i$'s except the mode $a_1$ of $N\mapsto 1$
% deleting machine before passing it on to  $1\mapsto M$ cloning machine.
Initially, after applying $N\mapsto 1$ deleting machine (\ref{delnm}) on the state $\ket{\psi}^{\otimes N}$ we will have 
$\rho_{a_1,a_2,...,a_N}^{del}$ as
%\begin{widetext}
\begin{eqnarray}
\rho_{a_1,..,a_N}^{del}=\alpha^{2N}\ket{N0}\bra{N0}
%\nonumber\\&+&
+\beta^{2N}\ket{1(N-1)0}\bra{1(N-1)0}
\nonumber\\ +\displaystyle\sum_{k=0}^{N-1}C(N-k,k)\alpha^{2(N-k)}\beta^{2k}
%\nonumber\\&&
\ket{(N-k)0,k1}\bra{(N-k)0,k1}.\nonumber\\
\label{delstate}
\end{eqnarray}
%\end{widetext}
%where $\binom{x}{y}=\frac{x!}{y! (x-y)!}$.
Then the reduced density 
matrix ($\rho_{a_i}^{del}$; $a_i\neq a_1$) of the state in equation (\ref{delstate}) is given by,
\begin{equation}
 \rho_{a_i}^{del}=\eta_0(\alpha,\beta,N)\ket{0}\bra{0}+\eta_1(\alpha,\beta,N)\ket{1}\bra{1},\label{reddelstate}
\end{equation}
where the form of the function $\eta_0$ and $\eta_1$ are
\begin{eqnarray}
 \eta_0(\alpha,\beta,N)&=&\alpha^{2N}+\displaystyle\sum_{i=1}^{N-1}\frac{(N-i)}{N} C(N,N-i)\alpha^{2(N-i)}\beta^{2i}\nonumber\\&+&\beta^{2N},\nonumber\\
 \eta_1(\alpha,\beta,N)&=&\displaystyle\sum_{i=1}^{N-1}\frac{i}{N}C(N,i)\alpha^{2(N-i)}\beta^{2i}.
\label{eta}
\end{eqnarray} 
 Now the state  $\rho_{a_i}^{del}$ (in equantion \ref{reddelstate}) is taken as input to $1\mapsto M$ cloning machine.
After the overall dual transformation the final reduced density matrix is,
\begin{widetext}
\begin{eqnarray}
% \ket{\psi}^{\otimes N}\mapsto 
\rho_f&=&\eta_0(\alpha,\beta,N)
%\nonumber\\&&
\displaystyle\sum_{j=0}^{M-1}\alpha_j^2\ket{(M-j)0,j1}\bra{(M-j)0,j1}
 \nonumber\\&+&
\eta_1(\alpha,\beta,N)\displaystyle\sum_{j=0}^{M-1}\alpha_{M-1-j}^2
%\nonumber\\&&
\ket{(M-1-j)0,(j+1)1}\bra{(M-1-j)0,(j+1)1}.
\end{eqnarray}
\end{widetext}
%$ ^nP_k=\frac{n!}{(n-k}!} - permutation \\  \binom nk=^nC_k=\frac{n!}{k!(n-k)!} - combination $ 
% where the parameter $\alpha_j$'s are
% \begin{eqnarray}
%  \alpha_j=\sqrt{\frac{2(M-j)}{M(M+1)}}.
% \end{eqnarray}
% Here we have considered the 2nd mode of the output state after deletion and then we fed that to BH cloning machine. This result is approximation of what we should do.
Finally, the reduced density matrix at first mode is given by,
% 2. With one particle conditional entropy.........
%\begin{widetext}
\begin{eqnarray}
\rho_f^a =\hspace{8cm}\nonumber\\
\displaystyle\sum_{j=0}^{M-1}\left\{\frac{(M-j)}{M} C(M,M-j)\eta_0
+\frac{j}{M}C(M,j)\eta_1)\right\}\alpha_j^2\ket{0}\bra{0}\nonumber\\ 
+\displaystyle\sum_{j=0}^{M-1}\left\{\frac{j}{M}C(M,j)\eta_0
+\frac{(M-j)}{M} C(M,M-j)\eta_1\right\}\alpha_{j}^2\ket{1}\bra{1}\nonumber
\end{eqnarray}
where $\eta_0$ and $\eta_1$ are given in equation (\ref{eta}).
Here also, we use bipartite quantum discord to quantify multiqubit quantum correlations in the dual physical process. 
The total correlations generated in this process is $\Delta_{\delta}^T=\Delta_{\delta}^{del}+\Delta_{\delta}^{clone}=\delta(\rho_f)$. 
In figure (\ref{1nnm1})  we once again have plotted the total correlations  $\Delta_{\delta}^T$ generated in the dual physical 
process of deletion followed by cloning against the state parameter $\alpha$ of the input state $ |\psi\rangle$. We have considered several 
cases and interestingly plots which show that the total correlations are always bounded. More precisely, the 
correlations generated in individual processes are complementary in nature,  i.e., 
$\Delta_{\delta}^{del}+\Delta_{\delta}^{clone}\leq 1$.
\section{Conclusions }
Complementarity is a fundamental feature of the quantum world which manifests in the dual physical nature of quantum particles. In this paper, we have shown a new kind of complementarity between two different physical processes such as approximate quantum cloning and  the deleting. We have shown that there is a relationship between quantum correlations generated in the process of 
cloning and deleting and the fidelity of the process in question. This has been illustrated using various measures of 
quantum correlations such as the geometric discord ($DG$), entropic quantum discord ($D$) and negativity ($N$). To bring out the generic nature of the complementarity, we 
have chosen three different classes of measure and irrespective of these measures we find that fidelity decreases with increase of 
correlations for both the processes of cloning and deletion.  This is well exhibited in terms of the amount of correlations generated 
in the successive processes of cloning and deletion (and vice versa). Moreover, we have witnessed an 
important property of quantum correlations called  
''complementarity" property in dual physical processes.
We have shown that the total correlations change in the cloning and the deleting is bounded by 
the maximum value of the measure of quantum correlations. We have illustrated  complementarity for a particular choice of 
cloning and deleting machine as well as for a particular measure  of correlations. We believe that this phenomenon is true for all classes of 
correlation measures and is independent of the 
choice of measure. It will be interesting to see if other quantum correlations display some complementary behavior 
in dual physical processes.

\noindent\textit{Acknowledgment:} Authors acknowledge Dr. S. Adhikari for having useful discussions.  
We would like to thank Dr. Giridhari Rao for going through our manuscript. We acknowledge the anonymous referee for useful suggestions which has improved the work immensely.

\end{document}